# Melting of tantalum at high pressure determined by angle dispersive x-ray diffraction in a double-sided laser-heated diamond-anvil cell

D. Errandonea[1, 2, †], M. Somayazulu[1], D. Häusermann[1], and H. K. Mao[3]

[1]HPCAT, Carnegie Institution of Washington, Advanced Photon Source, Building 434E, Argonne National Laboratory, 9700 South Cass Ave., Argonne, IL 60439, U.S.A.

[2]Departamento de Física Aplicada-ICMUV, Universitat de València, Edificio de Investigación, C/Dr. Moliner 50, 46100 Burjassot (Valencia), Spain

[3]Geophysical Laboratory, Carnegie Institution of Washington, 5251 Broad Branch Road, NW Washington, DC 20015, U.S.A.

The high pressure and high temperature phase diagram of Ta has been studied in a laser-heated diamond-anvil cell (DAC) using x-ray diffraction measurements up to 52 GPa and 3800 K. The melting was observed at nine different pressures, being the melting temperature in good agreement with previous laser-heated DAC experiments, but in contradiction with several theoretical calculations and previous piston-cylinder apparatus experiments. A small slope for the melting curve of Ta is estimated ($dT_m/dP$ ~ 24 K/GPa at 1 bar) and a possible explanation for this behaviour is given. Finally, a P-V-T equation of states is obtained, being the temperature dependence of the thermal expansion coefficient and the bulk modulus estimated.

PACS numbers: 62.50.+p, 64.70.Dv, 71.20.Be, 61.10.Nz

[†] Corresponding author, email: daniel.errandonea@uv.es , Te.: (34) 96 3543432, Fax: (34) 96 3543146

## 1. Introduction

Melting properties at high pressure (P) are of particular importance for understanding of the solid-liquid phase transition in high-pressure physics, material sciences, and geophysics. Specially, the study of melting under compression is important for improving the knowledge of the interior of planets and other celestial bodies, since most of their interiors are in a liquid state under high pressure. In recent years, the amount of available experimental data on melting at high pressure has increased considerably [1- 13]. As is known, melting at high pressure can be measured mainly by means of *in situ* laser-heated diamond-anvil cells (DAC) [1 – 12] and through shock-wave experiments [14 – 16]. In addition, many theoretical calculations [17 – 23] and empirical laws [24 – 27] have been developed to predict the melting curve of different materials under extreme compression. However, still today, these different methods yield widely different results.

Tantalum (Ta), one of the metals with the highest melting temperature ($T_M$) at ambient pressure (3269 K), is an excellent candidate to try to understand the melting properties at high pressure and because of this it has attracted a lot of attention [4, 16, 17, 18, 28]. At room temperature (RT), Ta remains stable in the simple body-centered cubic (bcc) structure up to about 180 GPa [29, 30] according to DAC experiments and is predicted to be stable in the bcc structure up to 1000 GPa (100 GPa = 1 Mbar) according to total-energy calculations [31]. Furthermore, at ambient pressure, the bcc phase of Ta is stable up to melting [32]. This high-structural stability of Ta allows the study of its pressure – volume – temperature (P-V-T) phase diagram, and particularly the effects of pressure on melting of Ta, over a wide compression range without the complication of structural changes. However, despite the experimental and theoretical efforts a consensus about the melting properties of Ta under pressure has not yet been reached even at low pressure. Fig. 1 illustrates the melting curves reported by different authors [4, 19, 28] showing that at 25 GPa a melting temperature of 3500 ± 100 K was measured using the *in situ* speckle method [4] whereas a melting temperature of 4400 K was calculated [19]. In Fig. 1 it can be also seen that the disagreements even include the slope of the melting curve ($dT_m/dP$) at 1 bar (different estimates give a large dispersion of values, e.g. 20 K/GPa [4], 53 K/GPa [28], and 98 K/GPa [18]). On the other hand, ultra high-pressure shock experiments identified the melting of Ta in the pressure range from 2.5 Mbar to 3 Mbar with $T_m$ estimated to be 7000 K – 1000K [16], while the extrapolation of the DAC data measured up to 1 Mbar [4] gives $T_m$ ~ 4500 K in the same pressure range. Disagreements resulting from different experimental and theoretical techniques are under debate and emphasize the need of additional studies.

Recently, angle-dispersive x-ray diffraction (ADXRD) in an externally heated DAC was employed successfully to determine melting of metals below 600 K [7]. ADXRD has also been recently combined with the laser-heating technique to perform structural studies of iron in the megabar pressure range at temperatures beyond 3000 K [11]. In this paper, we report a study of Ta using the same x-ray diffraction method in a double-sided laser-heated DAC up to a pressure of 52 GPa and a temperature (T) near to 3800 K. The melting temperatures determined at nine different pressures agree well with previous determinations obtained using the speckle method in a single-sided laser-heated DAC [4]. A possible explanation for the small slope for the melting curve of Ta found in the present and previous studies [4] is given. In addition, the experimental P-V-T data set collected is used to determine a high-temperature equation of states (EOS) for Ta.

## 2. Experimental Details

Commercial samples (Alfa Aesar) of stated purity 99.9 % were used to perform the studies reported here. Angle-dispersive x-ray diffraction measurements of Ta under high pressure and high temperature were performed in a double-sided laser-heated symmetric DAC (with flat diamonds with culet sizes ranging from 300 to 500 μm) at the 16ID-B undulator beamline of the High Pressure Collaborative Access Team (HPCAT), Advanced Photon Source (APS). A double-crystal branching monochromator (equipped with water-cooled diamond (111) and silicon (220) crystals) was used to produce a monochromatic x-ray beam with a wavelength of either $\lambda = 0.3738$ Å or $\lambda = 0.3698$ Å. The energy resolution of the monochromator was $\Delta E/E \sim 10^3$. The monochromatic x-ray beam was focused down, using multilayer bimorph mirrors in a Kickpatrick-Baez configuration [33, 34], to 10 μm by 10 μm. Diffraction images were recorded during 10 sec. with a MarCCD and were integrated and corrected for distortions using the FIT2D software [35]. The sample-CCD detector distance was either $\sim$ 258 mm (when $\lambda = 0.3738$ Å) or 280 mm (when $\lambda = 0.3698$ Å).

Ta samples compressed from powder with a diameter of 30 - 50 μm and a thickness of approximately 5 μm were loaded in stainless steel (grade 301) or rhenium gaskets with a pressure chamber having dimensions 100 - 150 μm in diameter and 30 μm thick. During sample loading, the Ta samples were located at the center of the gasket hole avoiding bridging the gasket by the sample. Dry sodium chloride (NaCl) was used as pressure-transmitting medium, acting also as thermal insulator between the sample and the diamond anvils. The ruby fluorescence technique [36] was applied to measure the pressure in the DAC at RT from an unheated ruby chip placed at 10 μm distance from the sample. The diffraction lines of NaCl were also used to estimate the pressure at RT and the change induced by the temperature increase based upon Decker's equation of states [37]. At room temperature, the highest pressure difference observed between the pressures obtained from both methods was 0.6 GPa at 50.5 GPa (see table I). Indexing and refinements to obtain the lattice parameters were carried out using XRDA [38] and GSAS [39] programs. A typical x-ray diffraction spectrum at room temperature has at least three diffraction peaks associated with Ta.

Ta samples were *in situ* double-sided laser-heated using the laser-heating system available at the HPCAT consisting of two identical Nd:YLF lasers (Photonics GS40, 85W, $TEM_{01}$ mode, $\lambda$ = 1053 nm). These lasers provide a total output of 170 W, with a power stability > 99 %, a beam pointing instability < 50 µrad, and sufficient collimation to obtain a laser-heating hot spot with a minimum radial temperature gradient in the center area. A 30 - 40 µm diameter hot spot with a temperature gradient < 10 K/µm was achieved at about T = 3800 K. Fig. 2 shows a photo of the hot spot obtained in the Ta sample at P ~ 8.65 GPa and T = 3230 ± 100 K. The laser beams were focused onto the sample using two 77 mm focal length apochromatic objective lenses (US Laser 3437) in a similar configuration to that described in Ref. [40]. The same lenses are used to collect the thermal radiation from both sides of the sample.

The temperature of the heated samples was measured with an accuracy of ± 100 K. These measurements were carried out using an Inspectrum 300 spectrograph equipped with a thermoelectric-cooled back-illuminated Hamamatsu CCD (1024 x 250 pixels). A two-leg fiber optic bundle, coupled to the double-entrance slit of the spectrograph [41], allowed us to collect simultaneously the thermal radiation from both sides of the sample, while preserving the spatial resolution of the inputs. This way, we had the ability of measuring the temperature of both sides of the sample at the same time using two different strips of the same CCD [41]. The size of each of the two chosen entrance slits of the spectrograph was 50 μm and the diameter of each individual optical fiber (which also acts as a pinhole) was also 50 μm, which is equivalent to about 4 μm on the sample. These measurements were performed in the spectral range 550-800 nm and the exposure time changed depending on the temperature from 0.5 to 3 sec. The main difference between our experimental set up and that reported in Ref. [40] is that our focusing optics are attached to the sample stage. Therefore, when moving the sample, in order to position the x-ray beam at its center, the hot spot remains fix on it. We estimated the x-ray beam and the lasers from both sides to be coincident within 3 μm. Then, the 30 – 40 μm hot spot fully covers the 10 μm size x-ray beam. In addition, a dual imaging set up with two CCD cameras (Panasonic WV-CP474) allowed us to visually observe the sample during the heating process and to check that the hot spot did not drift from the center of the sample.

Temperatures were determined by fitting the thermal radiation to the Planck radiation function [42]. The system response was calibrated using a standard tungsten ribbon lamp (OL550, Optronic Laboratories). Figure 3 shows a typical radiation spectrum after normalization to the system response and its fit to the Planck radiation function. The temperature difference between both sides of the sample was < 100 K. In order to know the temperature distribution on the sample, we used of a bundle of seven optical fibers which allowed us to simultaneously measure the temperature from seven different points of the same side of the sample using seven strips on the CCD of the Inspectrum 300 spectrograph [41]. One of the optical fibers collected the thermal radiation from the center of the hot spot and the remaining six the thermal radiation from the same number of points uniformly distributed in a circle of 30 μm diameter around the center of the hot spot (see Fig. 4(a)). Knowing T in these seven points we estimated the typical temperature distribution on a hot sample. Figure 4(a) shows the temperature simultaneously measured in seven different points of the Ta sample at P = 15.1 GPa and T = 3250 ± 100 K and the estimated temperature distribution reconstructed considering a radial temperature distribution as the one reported in Refs. [43] and [44] by using a B-spline approximation [45]. Figure 4(b) shows the obtained radial temperature profile in the same sample in those directions that minimize (y-axis of Fig. 4(a)) and maximize (x-axis of Fig. 4(a)) the temperature gradient.

## 3. Results and Discussions

### 3.1 Melting curve of Ta

The P-T behaviour of Ta was studied compressing the samples up to a desired pressure value at RT and then heating the sample at constant load up to temperatures where the Ta peaks disappeared. We followed this procedure in several samples at nine different pressures. In each case, the P-T path followed the sequence given in Table I. When cooling the sample to RT, no substantial change of the pressure was observed with respect to that measured before heating the sample (see Table I).

Figure 5 illustrates the typical differences in the diffraction pattern as the temperature increases in a sample pressurized up to 7.5 GPa at RT, being all the peaks arising from Ta and NaCl easily identified. This patterns allow the observation of the typical broadening and intensity reduction of the diffraction lines as the temperature increases [46]. The two main diffraction lines of the bcc phase of Ta, (111) and (200), were present at ~ 8.6 GPa and 3230 K, indicating that Ta was still crystalline. However, as the temperature was increased to 3450 K, we observed the disappearance of all diffraction lines and the appearance of some diffuse broad scattering (depicted by arrows in the upper trace of Fig. 5) together with a substantial increase of the background. We interpreted these facts as the onset of melting in Ta. This criterion has been successfully used before to the determine the melting of many different elements such as Fe [11], In [3], Kr [47], and Mg [48]. Upon cooling to RT, at all pressures, the diffraction lines of the bcc phase Ta were recovered (see Fig. 6), discarding the possibility that the diffraction peak disappearing could have been related to any chemical decomposition of the sample. Figure 6 shows diffraction patterns of Ta at several P-T conditions in a different sample pressurized up to 27.9 GPa at RT. In this case, the disappearance of the diffraction lines, the increase of the background and the appearance of the diffuse scattering characteristic of the onset of the melting occurs at 29.3 GPa and 3625 K. It is important to notice, that in all the studied samples the bcc phase of Ta was stable up to the onset of melting (i.e. there is not a phase of Ta different from bcc stable in the P-T range of this study) in good agreement with the previous established P-T phase diagram of Ta [32], which was constructed from high-pressure studies at RT, high-pressure melting experiments, high-temperature studies at 1 bar, and shock-wave experiments.

It has been shown that pressure increases in a laser-heated spot in a DAC [49 – 51], being this fact known as thermal pressure. The thermal pressure depends on the thermal expansivity and the compressibility of both the sample and the pressure medium and can be estimated by using a high-temperature calibrant [51]. In our case, we used the shift of the NaCl diffraction peaks, when laser-heating the Ta sample, to estimate the pressure change induced by the temperature increase. This shift can be easily identified by looking to the two lower traces of Fig. 5. On the upper traces, NaCl peaks were not observed because the temperature was above the melting temperature of NaCl [52]. Because of this fact, our method only allows us to determine *in situ* the thermal pressure on the sample at temperatures lower than the melting temperature of NaCl [52]. As expected, if the volume of the sample is small with respect to the volume of the pressure medium [51] as in our experimental configuration, the obtained values do not exceed 0.8 GPa at 2500 K (see Table I), in good agreement with previous estimations [52 - 54]. Then, at temperatures above the melting of NaCl [52] we estimated the thermal pressure using Ref. [54]. The estimated thermal pressures at all the different P-T conditions of our experiments are given in Table I.

Fig. 1 shows the melting results for Ta observed by us (open circles). Our measurements agree well with the melting curve determined using the speckle method (solid circles)[4], but they are lower than the data determined by Fateeva and Vereschagin using a piston-cylinder apparatus (solid squares) [28]. In Ref. [28] temperatures were estimated from the intensity ratios of thermal radiation measured in two narrow spectral ranges with the assumption that the emitted radiation is that of a black body. This method introduces large uncertainties in the temperature determination, which could easily explain the differences between the data reported in Ref. [28] and our data. From our previous results [4] a 1 bar $dT_m/dP \sim (20 \pm 4)$ K/GPa can be estimated. From the present data we estimated a value of $dT_m/dP \sim (24 \pm 2)$ K/GPa for the slope of the melting curve of Ta at 1 bar. These two values are nearly three times smaller than those estimated from methods based on first-principle calculations [18]. This fact raises concerns on the validity of calculating the pressure dependence of the melting properties of metals using models based on parameters calculated at 1 bar pressure. In Fig. 1 we present our present results of the melting of Ta together with previous experimental results [4, 28], theoretical results [19] and estimates obtained using the Lindemann law [25]:

$$\frac{\partial \ln T_M}{\partial \ln V_M} = \frac{2}{3} - 2\boldsymbol{g} \, , \qquad (1)$$

where $\gamma = 1.7$ [55] is the Grüneisen parameter and $V_M$ is the molar volume (V) of the solid at the melting state. In Fig. 1, it can be seen that melting estimates of Ta based on the Lindemann equation are not compatible with the present experimental data at any pressure. This fact is not surprising since the Lindemann law is an empirical law based on earlier experimental investigations of simple gases at low pressures, but it casts some doubts on the correctness of using the Lindemann law to calculate the melting behaviour of transition metals under extreme P-T conditions [56].

The low rate increase of the melting temperature reported here for Ta and previously for other transition metals [4] can be understood discussing melting in terms of the generation of vacancies [17, 57]. Within this framework, and using the Clausius-Clapeyron equation [58] we calculated the melting curve of Ta. The Clausius-Clapeyron equation follows directly from the Gibbs equality [59] and is written as:

$$\frac{\partial \ln T_M}{\partial P} = \frac{\Delta V_M}{\Delta H_M},$$ (2)

where $\Delta V_M$ and $\Delta H_M$ are, respectively, the difference in molar volume and enthalpy of the solid and liquid coexistent phases at melting conditions. To integrate equation (2), assuming $\Delta V_M/\Delta H_M$ independent of temperature [52], we need to know $\Delta V_M$ and $\Delta H_M$ as a function of pressure. For the pressure dependence of $\Delta H_M$ we assumed that is proportional to the vacancy formation enthalpy of Ta calculated by Mukherjee *et al*. [17] and for the pressure dependence $\Delta V_M$ we considered that is proportional to that calculated for iron by Alfe *et al*. [20]. Note that this latter pressure dependence goes like $P^{-4/5}$, as expected for transition metals [60]. In addition, as the ambient pressure values of $\Delta V_M$ and $\Delta H_M$ are unknown for Ta, we assume for this parameters the values reported for molybdenum ($\Delta H_M = 40.3$ KJ mol$^{-1}$ and $\Delta V_M = 0.3$ cm$^3$ mol$^{-1}$) [61], another transition metal whose melting curve follows the same trend than that of Ta [4]. The results obtained for the pressure dependence of $T_M$, $\Delta H_M$, and $\Delta V_M$ are shown in Fig. 1, Fig. 7(a), and Fig. 7(b), respectively. In Fig. 1, it can be seen that the present model reproduces well the trend of the experimental results at every pressure in spite of its simplicity, giving at ambient pressure a slope for the melting curve ($dT_m/dP \sim 24$ K/GPa) very close to the one observed by us.

Extrapolating the present results up to 3 Mbar a melting temperature of 4800 ± 300 K is obtained. This temperature is way below the one determined in shock-wave experiments ($T_M > 7000$ K) [16]. Direct temperature measurements in shock experiments require assumptions on the thermal and optical properties of the window material through which the sample is observed and the uncertainties may be of the order of 1000 K. Another issue is the superheating effects due too the small time scale of the shock experiments, which can lead to a 2000 K overestimation of $T_M$ [62, 63]. These two facts could probably explain the differences between our prediction and the shock-wave data [16]. Another question unanswered is whether there may be another factor at play in Ta, as the existence above 1 Mbar of a high pressure and high temperature phase in Ta like the one proposed for Mo [64]. This scenario will imply the existence of triple point at the P-T conditions where the solid-solid boundary line intercepts the melting curve. Usually, such a triple point would produce a discontinuous change in the slope of the melting curve [1, 2, 6], which could make converge the data measured below 1 Mbar [4] and the present calculations with the ultra high-pressure shock-wave data [16]. Clearly, a definitive understanding of the Ta phase diagram requires the extension of the laser-heating x-ray diffraction measurements here reported up to megabar pressures.

### 3.2 P-V-T equation of state

The RT compression data of Ta obtained from the experiments here reported (see Table 1 for a complete summary) are plotted in Figure 8 together with previous results [29, 30, 65, 66]. These data agree, within mutual experimental uncertainties, quite well with data reported in previous experiments [29, 30, 65]. However, below 15 GPa the data from the pioneering work of Ming *et al.* [66] show slightly higher volumes than those reported here. This is not surprising due to the higher resolution of our experiments, related to the fact that nowadays one can take advantage of the high instrumental resolution of the area detectors (the MarCCD in our case) and of the high brilliance reached at the APS. A Birch – Murnaghan third order EOS [67] fitted to our data and those reported by Cynn *et al.* [29] and Hanfland *et al.* [30] yield the following parameters for RT the bulk modulus, its pressure derivative, and the molar volume of Ta at ambient conditions, respectively: $B_0 = (190 \pm 15)$ GPa, $B_0^{'} = 3.7 \pm 0.5$, and $V_0 = 10.85 \pm 0.08$ cm$^3$/mol. These parameters are in good agreement with those previously reported [29, 30, 65, 66], which average $B_0 = (198 \pm 10)$ GPa, $B_0^{'} = 3.4 \pm 0.5$, and $V_0 = (10.90 \pm 0.15)$ cm$^3$/mol.

Finally, we used all the data from this study (see Table I), previous RT compression data [29, 30], and previous ambient pressure high-temperature data [68] to obtain a P-V-T relation for Ta. For this purpose we used the Birch-Murnaghan isothermal formalism [66, 69]:

$$P(V,T) = \frac{3}{2} B_{0,T} \left( \left( \frac{V_{0,T}}{V} \right)^{7/3} - \left( \frac{V_{0,T}}{V} \right)^{5/3} \right) \times \left( 1 + \frac{3}{4} \left( B_{0,T}^{'} - 4 \right) \left( \left( \frac{V_{0,T}}{V} \right)^{2/3} - 1 \right) \right) \quad , \quad (3)$$

where

$$V_{0,T} = V_0 \exp\left(\int\limits_{300\,K}^{T} \boldsymbol{a}\left(T\right)\partial T\right) \quad,\qquad\qquad (4)$$

and

$$B_{0,T} = B_0 + \frac{\partial B_{0,T}}{\partial T}\left(T - 300\,K\right) \quad.\qquad\qquad (5)$$

Since usually $\dfrac{\partial B_{0,T}}{\partial T} \gg \dfrac{\partial B_{0,T}^{'}}{\partial T}$ [47, 69], we assumed $B_{0,T}^{'} = B_0^{'}$. In addition, a

linear behavior of the thermal expansion coefficient was assumed, with

$\boldsymbol{a} = \boldsymbol{a}_0 + \dfrac{\partial \boldsymbol{a}}{\partial T}(T - 300\,K)$, where $\boldsymbol{a}_0 = 20 \times 10^{-6}\,K^{-1}$ is the thermal expansion at 300 K

[68]. With these assumptions, $\dfrac{\partial \boldsymbol{a}}{\partial T}$ and $\dfrac{\partial B_{0,T}}{\partial T}$ are the only two parameters to be

determined. By fixing $B_0$ = 190 GPa, $B_0^{'}$ = 3.7, and $V_0$ = 10.85 cm³/mol, we obtained

$\dfrac{\partial \boldsymbol{a}}{\partial T} = (1.6 \pm 0.5)\ 10^{-9}\,K^{-2}$ and $\dfrac{\partial B_{0,T}}{\partial T} = (-1.8 \pm 0.3)\ 10^{-2}\,GPa\ K^{-1}$. These values

compare well with those previously measured [68] and calculated [70] for Ta and indicate a decrease of the bulk modulus and an increase of the thermal expansion with temperature.

## 4. Summary

We studied the melting and the structural properties of Ta under pressure in a

DAC up to 52 GPa and 3800 K combining the use of a micro x-ray beam and the

laser-heating technique. The sharp x-ray distribution and the homogeneous

temperature distribution achieved are critical for high P-T x-ray diffraction

experiments in order to obtain quality data. The obtained results confirm previous

DAC results experiments that were in conflict with theoretical calculations and earlier piston-cylinder experiments. We observed that the melting slope of Ta is small, being its value $dT_m/dP \sim$ 24 K/GPa at atmospheric pressure. Interpreting the melting in terms of the generation of vacancies we provide a plausible explanation for the experimentally observed behavior. Furthermore, a P-V-T- relation for Ta is presented which describes well the present data and those found in the literature [29, 30, 68]. The temperature dependence of the bulk modulus and the thermal expansion were estimated from the present data, being the obtained values for their temperature derivatives $\frac{\partial a}{\partial T} = (1.6 \pm 0.5) \ 10^{-9} \ K^{-2}$ and $\frac{\partial B_{0,T}}{\partial T} = (-1.8 \pm 0.3) \ 10^{-2} \ GPa \ K^{-1}$.

### Acknowledgments

This work was supported by the NSF, the DOE, and the W. M. Keck Foundation. We would like to thank the rest of the staff at the HPCAT at APS for their help to set up the experimental system and for their contribution to the success of the experiments. Daniel Errandonea acknowledges the financial support from the MCYT of Spain and the Universitat of València through the "Ramón y Cajal" program for young scientists. He is also particularly grateful to Russell Jurgensen from Effective Objects for its help in developing the temperature measurements software.

**Figure Captions**

**Figure 1:** Melting curve of Ta. Experimental data: (?) present work, (?) Ref. [4], and (¦) Ref. [28]. The dot-dashed line illustrates the melting curve calculated in Ref. [19]. The double dot-dashed line shows the Lidenmann law's estimates. The dashed line fits the present melting data and the dotted one those reported in Ref. [4]. The solid line is melting curve calculated using the present model.

**Figure 2:** 50 µm Ta sample laser-heated in a DAC with a stainless steel gasket with a hole of 150 µm diameter. P = 8.65 GPa and T = 3230 K.

**Figure 3:** Example of the thermal radiation normalized to the system response and the fit (dotted line) to the Planck radiation function. The shown spectrum (exposure time 1 sec.) corresponds to a temperature of 2630 K in Ta sample at 40.4 GPa.

**Figure 4: (a)** Estimated temperature distribution in a Ta sample at P = 15.1 GPa. The central temperature of the hot spot is 3250 K. The external circle represents the size of the hot spot. The small solid circles are the areas from where the temperature was measured (Black = 3250 K, dark gray = 3220 K, and gray = 3200 K). The dotted lines are the estimated temperature contour lines. **(b)** Temperature gradient of the same sample in those directions that minimize (dotted line) and maximize it (solid line).

**Figure 5:** X-ray diffraction pattern at different temperatures in a sample compressed up to 7.5 GPa at RT. Temperature and pressure of every pattern are indicated in the figure. Ta and NaCl diffraction peaks are identified in the lower trace. The background was substracted. The arrows depict the broad scattering characteristec of the onset of melting. These experiment were performed at $\lambda$ = 0.3738 Å and the detector to sample distance was ~ 258 mm.

**Figure 6:** X-ray diffraction pattern at different temperatures in a sample compressed up to 27.9 GPa at RT. Temperature and pressure of every pattern are indicated in the figure. Ta and NaCl diffraction peaks are identified in the lower trace. The upper trace shows a diffraction pattern obtained upon cooling to RT after melting. The background was substracted. The arrows depict the broad scattering characteristec of the onset of melting. These experiment were performed at $\lambda = 0.3698$ Å and the detector to sample distance was ~ 280 mm.

**Figure 7:(a)** Assumed pressure dependence of the enthalpy change at melting. **(b)** Assumed pressure dependence of the volume change at melting.

**Figure 8:** RT P-V data of Ta. (?) present data, (?)Ref. [66], (?) Ref. 29, (?) Ref. 30. The diamond with the error bar indicates the average and the standard deviation of the value reported in the literature [29, 30, 65, 66] for the molar volume at ambient conditions. The solid line represent the fitted P-V relation.

**Table Captions**

**Table I:** Unit-cell parameter and volume of bcc Ta at different P-T conditions. The estimated errors are 0.1% and 0.3% respectively. The estimated thermal pressure and the observance of melting are indicated.

**Table I**

| Temperature [K] | RT Ruby Pressure [GPa] | NaCl Pressure [GPa] | Thermal Pressure [GPa] | Sample Pressure [GPa] | Phase | $a$ [Å] | V [cm$^3$/mol] |
|---|---|---|---|---|---|---|---|
| 300 | 2.5 | 2.75 | | 2.75 | bcc | 3.276 | 10.590 |
| 1250 | | 3.05 | 0.3 | 3.05 | bcc | 3.287 | 10.702 |
| 2270 | | | 0.7 | 3.45 | bcc | 3.299 | 10.818 |
| 3190 | | | 1.15 | 3.9 | bcc | 3.306 | 10.885 |
| 3350 | | | 1.25 | 4 | liquid | | |
| 300 | 2.5 | 2.7 | | 2.7 | bcc | 3.274 | 10.576 |
| | | | | | | | |
| 300 | 7.45 | 7.5 | | 7.5 | bcc | 3.248 | 10.325 |
| 1610 | | 7.9 | 0.4 | 7.9 | bcc | 3.270 | 10.536 |
| 2450 | | | 0.75 | 8.45 | bcc | 3.288 | 10.705 |
| 3230 | | | 1.15 | 8.65 | bcc | 3.304 | 10.869 |
| 3450 | | | 1.3 | 8.8 | liquid | | |
| 300 | 5.7 | 5.85 | | 5.85 | bcc | 3.278 | 10.613 |
| | | | | | | | |
| 300 | 13.9 | 13.9 | | 13.9 | bcc | 3.227 | 10.119 |
| 1480 | | 14.3 | 0.4 | 14.3 | bcc | 3.246 | 10.304 |
| 2520 | | | 0.8 | 14.7 | bcc | 3.263 | 10.463 |
| 3250 | | | 1.2 | 15.1 | bcc | 3.274 | 10.568 |
| 3515 | | | 1.35 | 15.25 | liquid | | 10.604 |
| 300 | 13.6 | 13.8 | | 13.8 | bcc | 3.231 | 10.159 |
| | | | | | | | |
| 300 | 15 | 14.95 | | 14.95 | bcc | 3.231 | 10.159 |
| 2210 | | | 0.65 | 15.6 | bcc | 3.254 | 10.377 |
| 2600 | | | 0.85 | 15.8 | bcc | 3.258 | 10.415 |
| 3380 | | | 1.25 | 16.2 | bcc | 3.265 | 10.479 |
| 3505 | | | 1.35 | 16.3 | liquid | | |
| 300 | 14.8 | 14.75 | | 14.75 | bcc | 3.232 | 10.168 |
| | | | | | | | |
| 300 | 18.7 | 19 | | 19 | bcc | 3.214 | 9.993 |
| 1630 | | 19.4 | 0.4 | 19.4 | bcc | 3.227 | 10.119 |
| 2085 | | | 0.6 | 19.6 | bcc | 3.232 | 10.168 |
| 2540 | | | 0.8 | 19.8 | bcc | 3.235 | 10.198 |
| 2885 | | | 1 | 20 | bcc | 3.239 | 10.220 |
| 3450 | | | 1.3 | 20.3 | bcc | 3.241 | 10.247 |
| 3560 | | | 1.35 | 20.35 | liquid | | |
| 300 | 18.2 | 18.4 | | 18.4 | bcc | 3.216 | 10.017 |
| | | | | | | | |
| 300 | 21.5 | 21.75 | | 21.75 | bcc | 3.202 | 9.887 |
| 1920 | | 22.3 | 0.55 | 22.3 | bcc | 3.217 | 10.022 |
| 2370 | | | 0.75 | 22.5 | bcc | 3.220 | 10.053 |
| 2935 | | | 1 | 22.75 | bcc | 3.223 | 10.085 |
| 3450 | | | 1.3 | 23.05 | bcc | 3.225 | 10.099 |
| 3585 | | | 1.4 | 23.15 | liquid | | |
| 300 | 21.1 | 21.35 | | 21.35 | bcc | 3.204 | 9.902 |
| 300 | 27.5 | 27.9 | | 27.9 | bcc | 3.178 | 9.667 |
| 1900 | | 28.45 | 0.55 | 28.45 | bcc | 3.189 | 9.764 |
| 2515 | | | 0.8 | 29.05 | bcc | 3.192 | 9.791 |
| 3140 | | | 1.15 | 29.15 | bcc | 3.193 | 9.801 |
| 3400 | | | 1.25 | 29.15 | bcc | 3.193 | 9.801 |
| 3520 | | | 1.35 | 29.25 | bcc | 3.192 | 9.799 |
| 3625 | | | 1.4 | 29.3 | liquid | | |
| 300 | 26.8 | 27.1 | | 27.1 | bcc | 3.181 | 9.694 |
| | | | | | | | |
| 300 | 39.3 | 39.6 | | 39.6 | bcc | 3.138 | 9.300 |
| 1500 | | 40 | 0.4 | 40 | bcc | 3.142 | 9.339 |
| 2630 | | 40.4 | 0.8 | 40.4 | bcc | 3.143 | 9.351 |
| 3025 | | | 1.05 | 40.65 | bcc | 3.142 | 9.339 |
| 3340 | | | 1.35 | 40.95 | bcc | 3.140 | 9.325 |
| 3520 | | | 1.40 | 41 | bcc | 3.139 | 9.317 |
| 3670 | | | 1.45 | 41.05 | liquid | | |
| 300 | 38.9 | 39.1 | | 39.1 | bcc | 3.139 | 9.137 |
| | | | | | | | |
| 300 | 49.9 | 50.5 | | 50.5 | bcc | 3.104 | 9.008 |
| 1920 | | 51.05 | 0.55 | 51.05 | bcc | 3.106 | 9.020 |
| 2480 | | 51.30 | 0.80 | 51.30 | bcc | 3.105 | 9.010 |
| 3025 | | | 1.05 | 51.55 | bcc | 3.102 | 8.988 |
| 3440 | | | 1.3 | 51.8 | bcc | 3.099 | 8.959 |
| 3635 | | | 1.4 | 51.9 | bcc | 3.097 | 8.943 |
| 3780 | | | 1.5 | 52 | liquid | | |
| 300 | 49.3 | 49.5 | | 49.5 | bcc | 3.107 | 9.033 |

**Figure 1:**

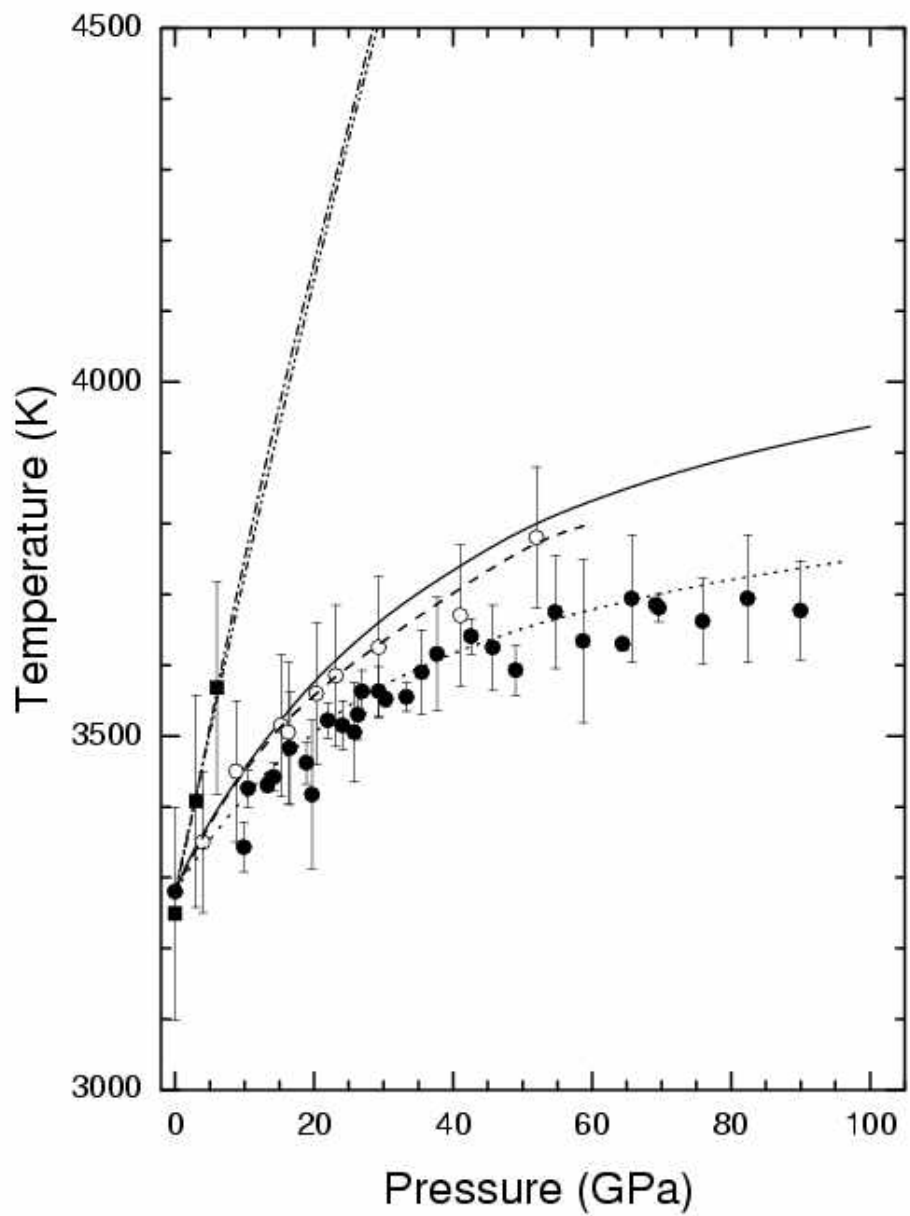

**Figure 2:**

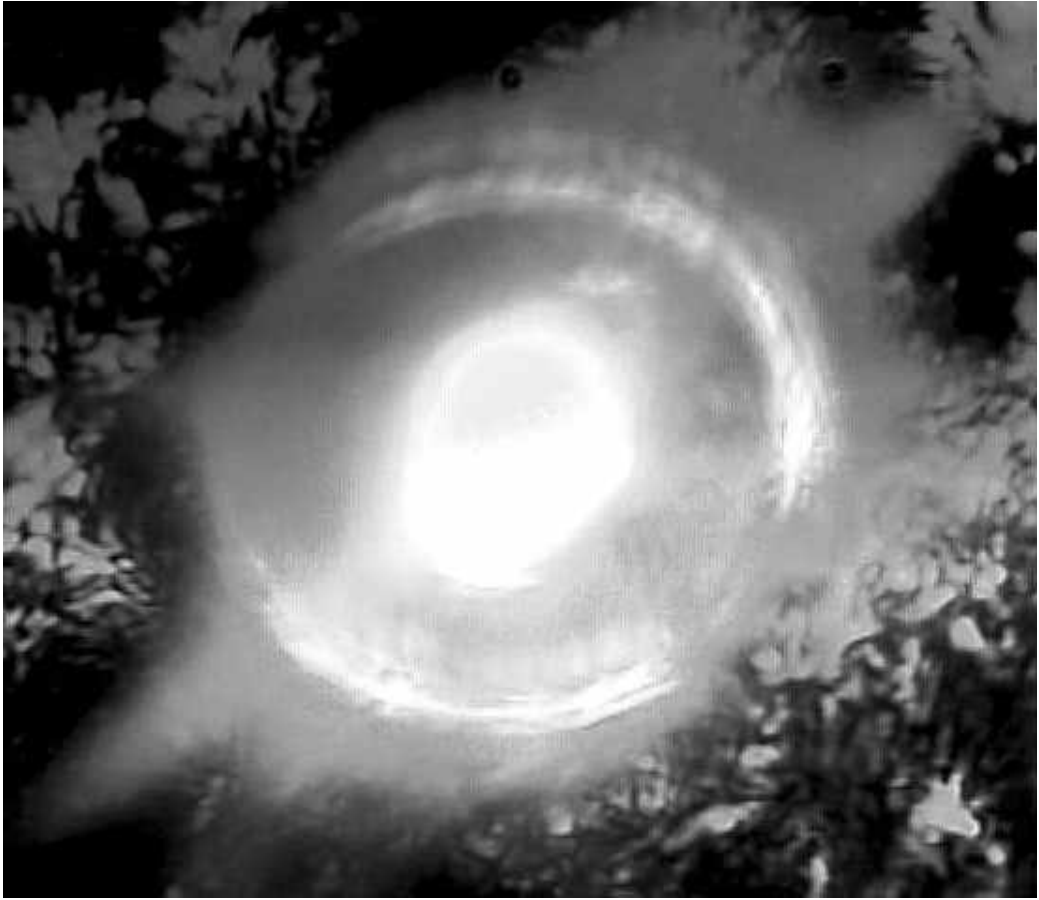

**Figure 3:**

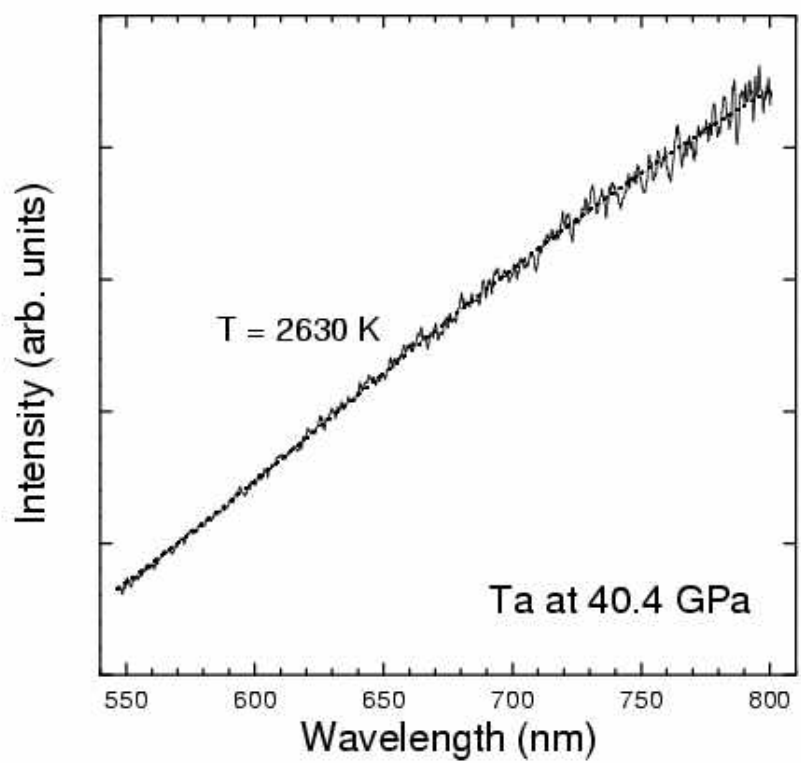

**Figure 4:**

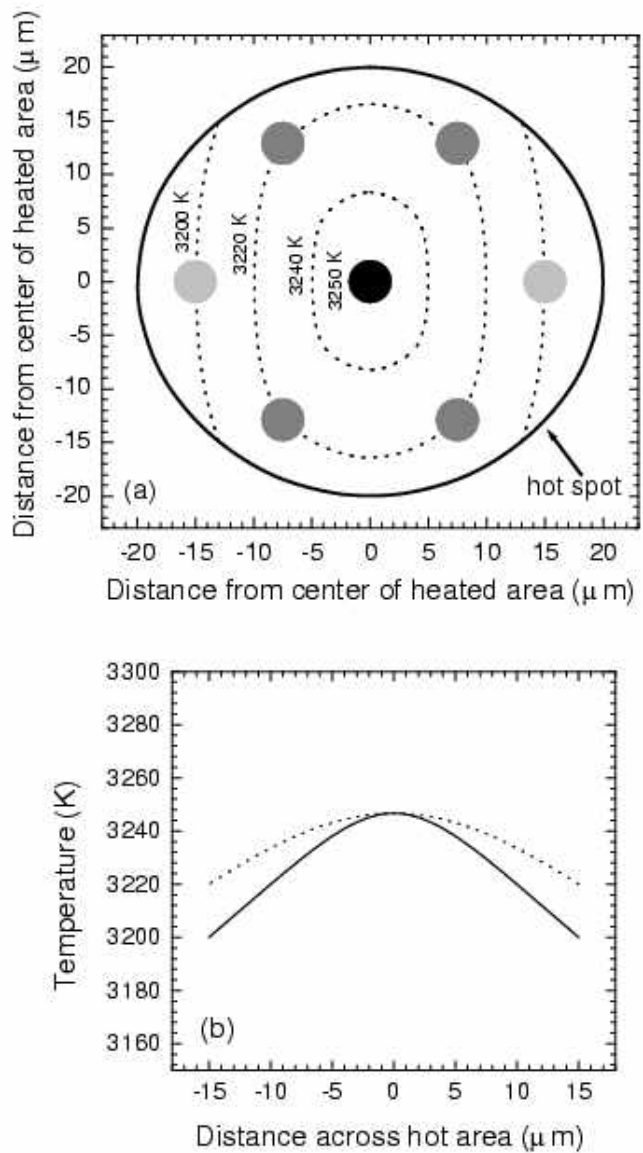

**Figure 5:**

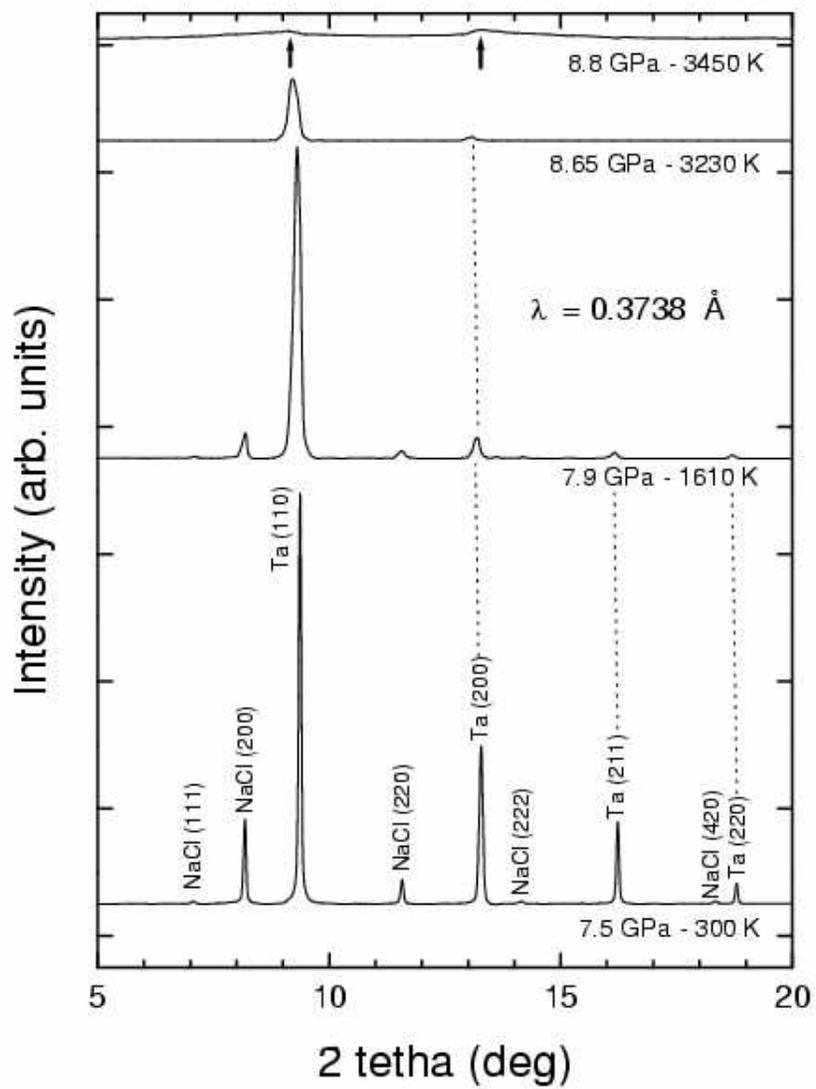

**Figure 6:**

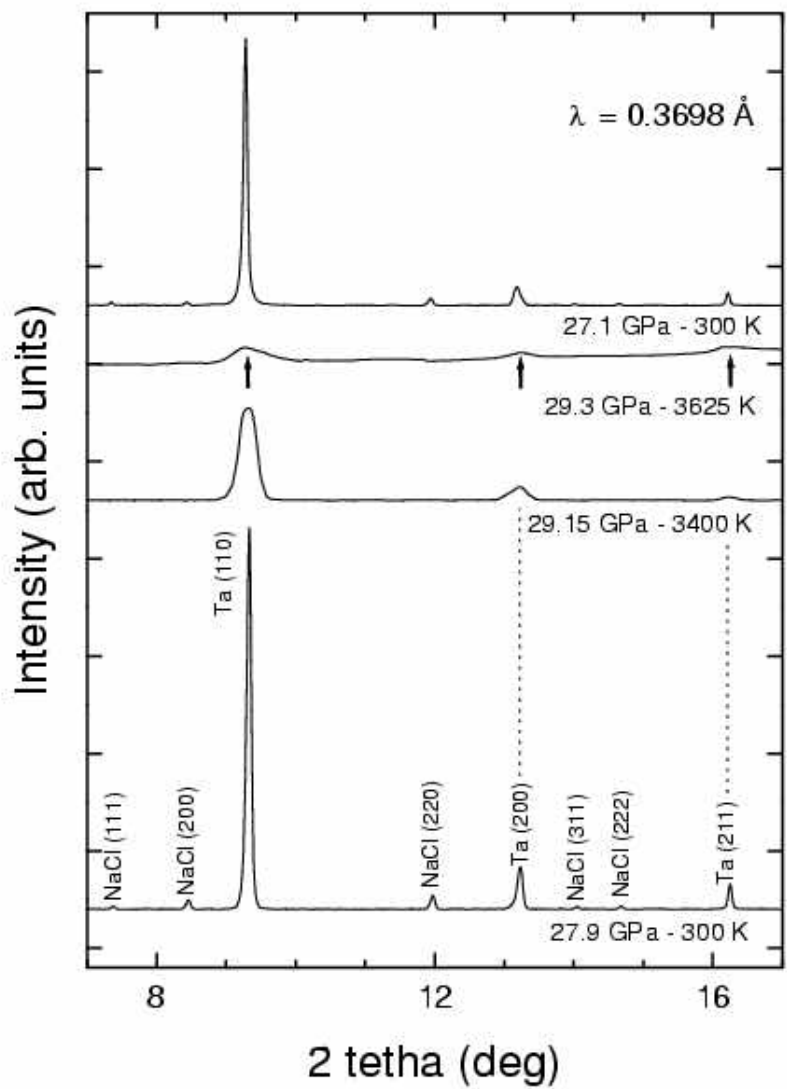

**Figure 7:**

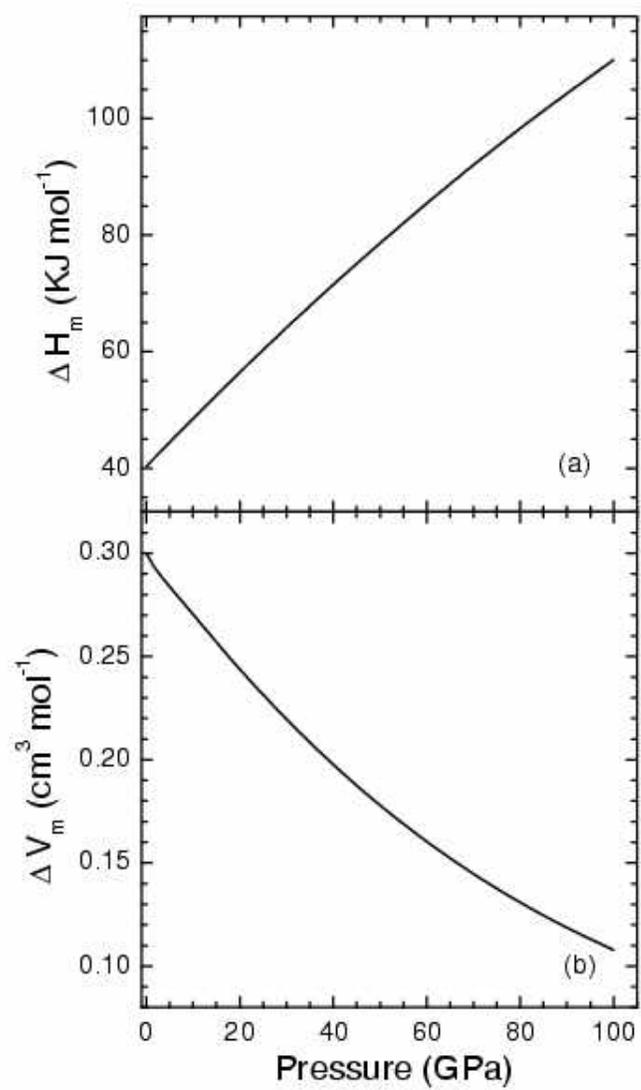

**Figure 8:**

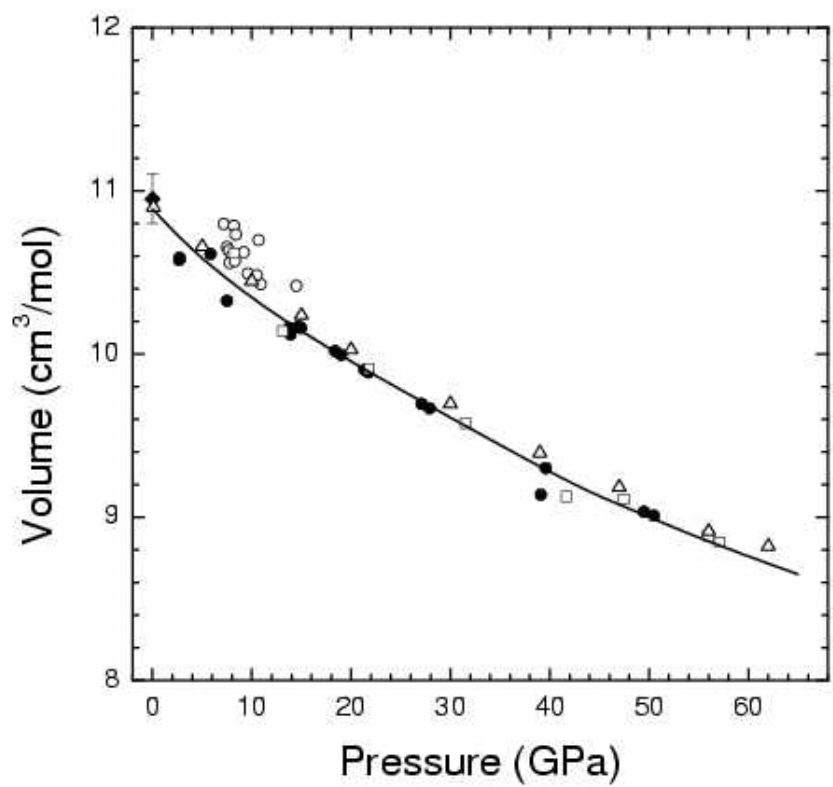